\documentclass[letterpaper, 10pt, conference]{ieeeconf}      
\IEEEoverridecommandlockouts                              
                                                          
\let\labelindent\relax
\usepackage{graphicx}   
\usepackage{epsfig} 
\usepackage[T1]{fontenc} 
\usepackage{amssymb} 
\usepackage{amsmath} 
\usepackage{graphicx}
\usepackage{lipsum}%
\usepackage{xcolor}
\usepackage{hyperref}
\usepackage{float}
\usepackage{enumitem}
\usepackage[backend=bibtex,style=ieee,sorting=none]{biblatex}
\bibliography{root.bib}
\title{\LARGE \bf Anomaly Detection from Cyber Threats via Infrastructure to Automated Vehicle} 

\author{Chris van der Ploeg$^{1,2}$, Robin Smit$^{1}$, Alexis Siagkris-Lekkos$^{1}$, Frank Benders$^{1}$, Emilia Silvas$^{1,3}$ 
\thanks{$^{1}$Netherlands Organisation for Applied Scientific Research, Integrated Vehicle Safety Group, 5700 AT Helmond, The Netherlands.}
\thanks{$^{2}$Eindhoven University of Technology, Dynamics and Control Group, Mechanical Engineering Dept., P.O. Box 513, 5600 MB, Eindhoven, The Netherlands. (e-mail: \href{mailto:c.j.v.d.ploeg@tue.nl}{c.j.v.d.ploeg@tue.nl}).}
\thanks{$^{3}$Eindhoven University of Technology, Control Systems Technology Group, Mechanical Engineering Dept., P.O. Box 513, 5600 MB, Eindhoven, The Netherlands.}
}

\begin{document}
\maketitle
\thispagestyle{empty}
\pagestyle{empty}
\begin{abstract}                
Using Infrastructure-to-Vehicle (I2V) information can be of great benefit when driving autonomously in high-density traffic situations with limited visibility,  since the sensing capabilities of the vehicle are enhanced by external sensors. In this research, a method is introduced to increase the vehicle's self-awareness in intersections for one of the largest foreseen challenges when using I2V communication: cyber security. 
The introduced anomaly detection algorithm, running on the automated vehicle, assesses the health of the I2V communication against multiple cyber security attacks. The analysis is done in a simulation environment, using cyber-attack scenarios from the Secredas Project (Cyber Security for Cross Domain Reliable Dependable Automated Systems) and provides insights into the limitations the vehicle has when facing I2V cyber attacks of different types and amplitudes and when sensor redundancy is lost. 
The results demonstrate that anomalies injected can be robustly detected and mitigated by the autonomous vehicle, allowing it to react more safely and comfortably and maintaining correct object tracking in intersections.
\end{abstract}

\section{INTRODUCTION}
Connected and automated driving is emerging as a solution to safer, more efficient, sustainable and  comfortable road transport system, \cite{Ellen2012}. Different driver assisting or autonomous functions are brought to the road by vehicle manufacturers, with increased connectivity possibilities, such as vehicle-to-vehicle (V2V), vehicle-to-infrastructure (V2I, I2V) or vehicle-to-everything (V2X, X2V). These  bring functionality and performance benefits but also risks and concerns in terms of cyber security and privacy: the automated vehicle is a \textit{thing} in the Internet of Things (IoT), that can cause dangerous situations and that collects large amounts of data. 

Via X2V communications it is now possible to breach a vehicle without being in a certain physical range or using previously installed hardware, \cite{Checkoway2011}. A lack of cyber security can lead to unsafe situations and will result in a lack of trust and reluctant users. This calls for improved safe- and secure-by-design concepts for connected and automated driving that can detect spoofed and tampered X2V messages and will react proactively and safely.

An anomaly, defined in the ISO 26262 standard as a condition that deviates from expectations, based for example on requirements, specifications, design documents, user documents, standards or on experience \cite{iso}, can have multiple forms and patterns. More specifically, cyber network anomalies (i.e. intrusion attempts or threats) are used to access and manipulate information or to render a system as unreliable or unusable. 
Detecting cyber network anomalies refers to the problem of detecting in real-time wrong patterns and anomalies in I2V traffic that do not conform to the expected normal behaviour,  \cite{Bhuyan2014}.
I2V communication can be very beneficial especially in intersection scenarios where a Road Side Unit (RSU) is used to communicate the position of observed Vulnerable Road Users (VRU), enhancing the tracking range and the state estimation accuracy of the detected objects, \cite{Grembek2019}.

\begin{figure}[h!]
    \centering
    \includegraphics[clip, trim=10mm 10mm 140mm 20mm, width=0.7\columnwidth]{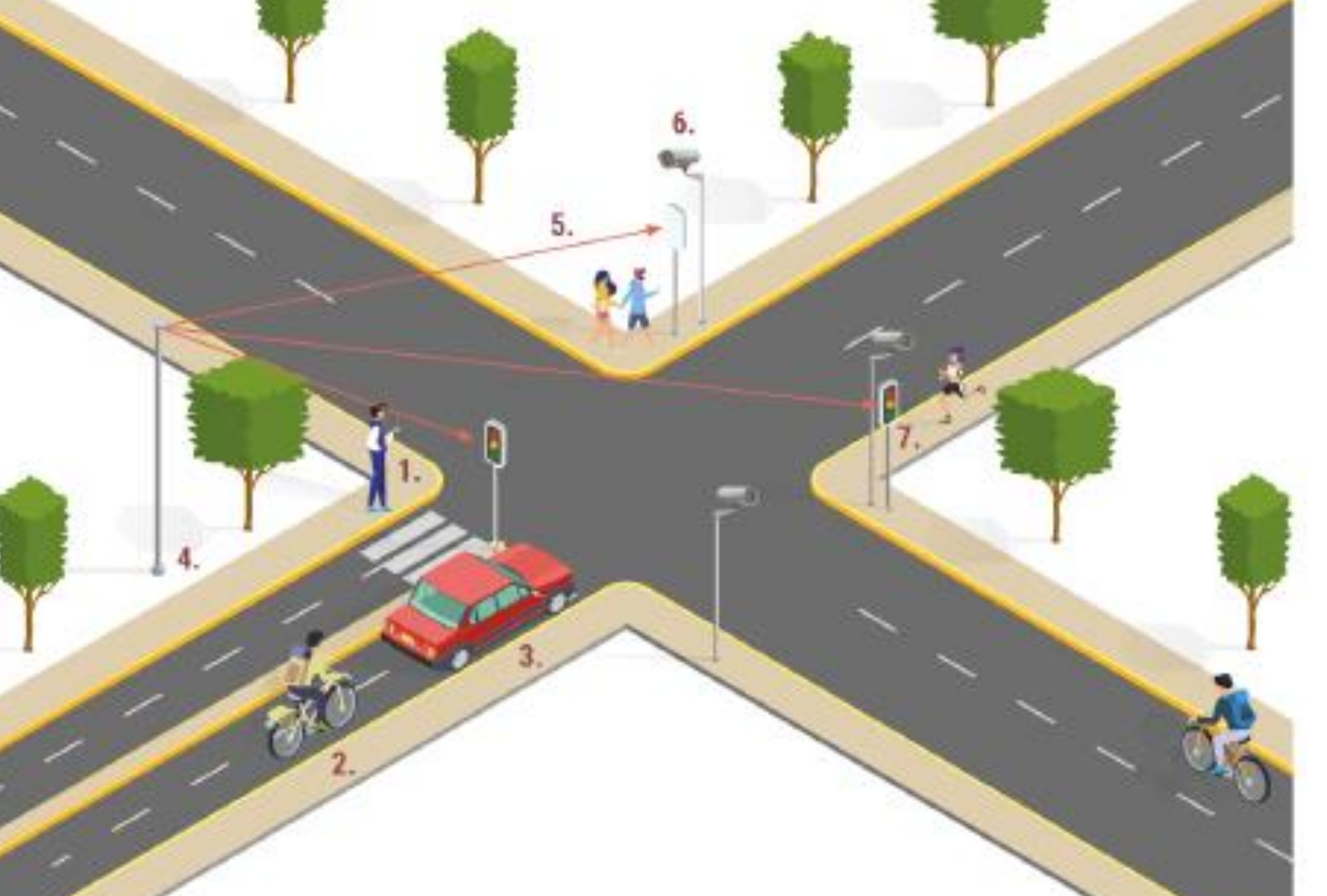}
    \caption{Intersection crossing scenario considered, where road side units (RSUs) communicate information to the vehicle about other road users.}
    \label{fig:scene}
\end{figure}

Multiple anomaly detection methods have been developed that can be applied for I2V cyber attacks as well, with many focusing on single and known types of anomalies for limited scenarios or on taking into account the state of another leading vehicle as available in cooperative driving, \cite{Wang2019}. For example, particle filtering and maximum likelihood methods are used in \cite{Lampiri2017}, comparing sensor outputs using pairwise inconsistency graphs in \cite{Park2015}, \cite{van_Wyk2019} uses a combination between a convolutional neural network and a Kalman filter and \cite{negi2020} uses long short-term memory (LSTM) neural networks. Furthermore, several studies have introduced the types of anomalies that can typically be expected from cyber attacks, \cite{Bhuyan2014,van_Wyk2019} and the possible attackers' behaviours, \cite{Panda2018}, yet no methods have been introduced that can deal with multiple types of attacks around intersections, via I2V communication, as depicted in Fig. \ref{fig:scene}. It is also unknown how sensor redundancy can help in ensuring a safe operation of the vehicle and can be used to identify the anomaly.

To address this challenge, in this paper we propose an extended Kalman filter (EKF)  to detect object tracking anomalies that result from in-vehicle cyber attacks coming from I2V communication. We chose this algorithm to benefit from the domain knowledge about the vulnerable road users and since EKF is a well-established anomaly detection
method, \cite{van_Wyk2019}. 
Unlike previous works, we develop and validate our method on multiple types of attacks for intersection scenarios that the automated vehicle can encounter and we discuss the importance of sensor redundancy depending on different anomalies. 

This paper is organized as follows. Section \ref{preliminaries} introduces the functional architecture of the automated vehicle and the types of attacks considered while Section \ref{sec:problemdescription} introduces the anomaly detection problem, the scenarios and the developed simulation environment. Sections \ref{algorithm} and \ref{results} introduce the extended Kalman filter and simulation results resulting from this design.  Finally, the conclusions and recommendations for future work are described in Section \ref{conc}.

\section{Preliminaries}\label{preliminaries}
 
 In an automated driving (AD) vehicle, internal and possibly external sensors are used to monitor the environment. The system architecture and the types of anomalies are important to be able to develop a robust method for detecting these anomalies. This section introduces the vehicle functional architecture used, an overview of the types of existing anomalies and a selection relevant for cyber-attacks via I2V.

\subsection{Sensor fusion and object tracking}
\begin{figure}[h!]
   \centering
    \includegraphics[clip, trim=38mm 20mm 62mm 20mm, width=0.9\columnwidth]{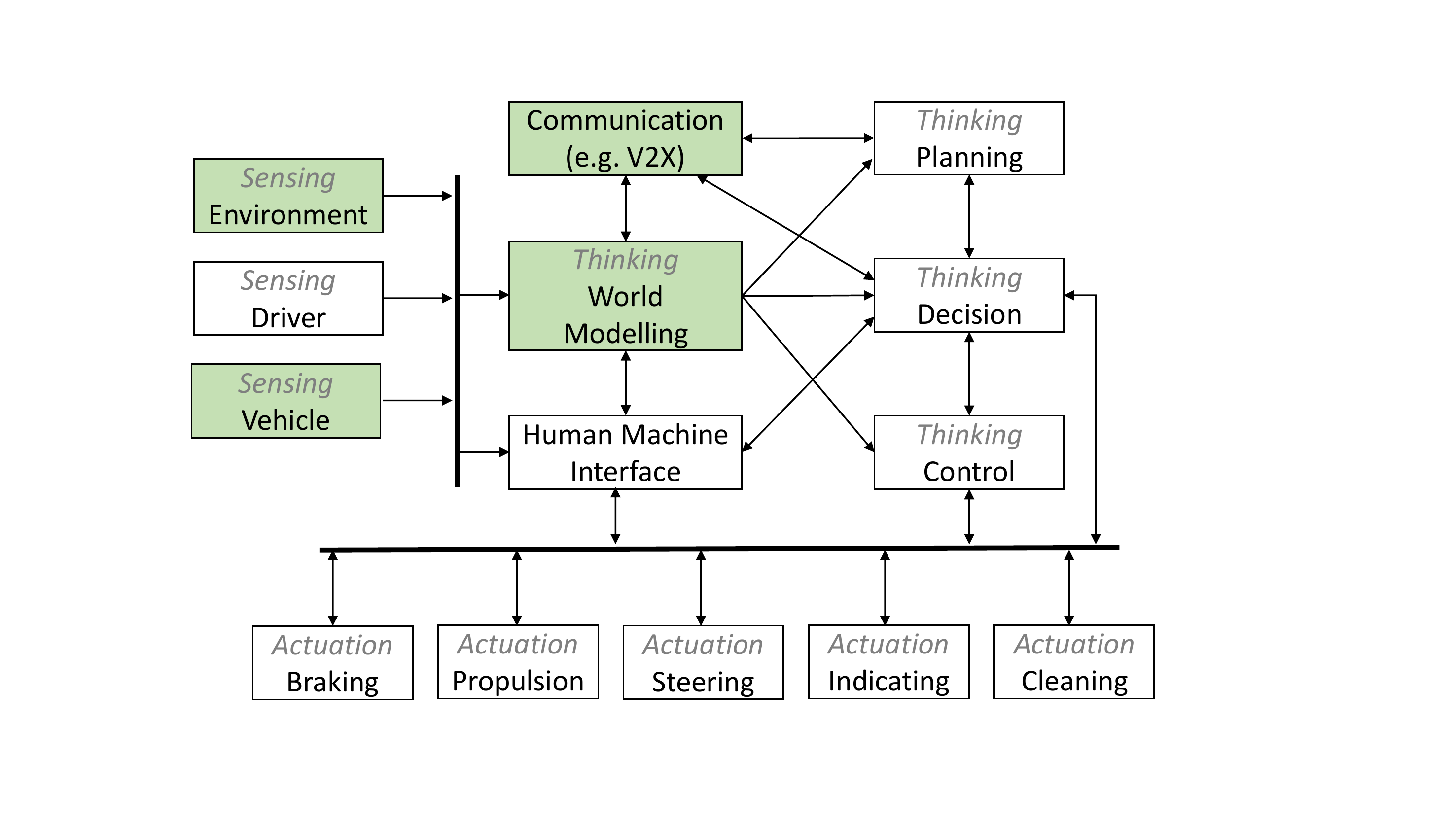}
    \caption{The functional architecture of the AD application, highlighting the used components for I2V anomaly detection (based on \cite{Arashsafety,arashphd})}
    \label{safety}
\end{figure}

The functional architecture of the AD vehicle is depicted in Fig. \ref{safety}, where world modeling contains the \textit{ego vehicle state estimation} functions, \textit{object detection and tracking} functions and \textit{road modeling} functions, such as \textit{lane} and \textit{empty drivable space} models.
The dynamic objects in the vehicle's environment are tracked by on-board sensors (i.e., radar, lidar, camera) providing relative measurements to the vehicle. Furthermore, in the scenario of Fig. \ref{fig:scene}, the camera attached to the RSU is providing absolute measurements using I2V communication. These sensor measurements, together with the ego vehicle state estimation, are fused into a single object state estimate by a module, i.e.  \textit{Target Tracker}.

\subsection{Types of anomalies}
Anomalous sensor behaviour can have different forms, which pose different challenges in detecting the deviation from the healthy behaviour. Besides a malicious attack, these anomalies could come from e.g. wrong sensor calibration, limited sensor capabilities and environmental influence, \cite{van_Wyk2019}.
Categories of possible sensor signals anomalies are introduced in \cite{Sharma2010}, with three main classes being explained as important and observed in real-life deployments, i.e. \textit{short}, \textit{noise} and \textit{constant}. In more detail, the authors in  \cite{van_Wyk2019} distinguish between four types of anomalies, i.e. \textit{bias}, \textit{gradual drift}, \textit{instant} and \textit{miss}. 
\textit{Bias} anomalies imply that the anomalous signal has a temporally constant error compared to the \textit{normal} sensor reading. 
\textit{Gradual drift} refers to a gradual drift (increasing or decreasing) in the observed data, for a period of time, with or without a bias. The longer the period of the drift, the more dangerous and unrepresentative the specific signal can become with respect to the true state of the observed object.
\textit{Instant} anomalies imply  sharp, unexplained changes in the observed data between two successive sensor readings, while \textit{miss} anomalies refer simply to the lack of available object or signal data during a time period. The authors in \cite{Wang2019} consider  that additional anomalies can be \textit{noise}, in which the signal gets perturbed (e.g. random variation of brightness or color information in an camera image), or \textit{burst}, where the signal gets disturbed for a short amount of time (e.g., driving over a pothole).

To ensure a safe operation of the ego vehicle in case of I2V cyber-attacks, see Fig. \ref{fig:scene}, it is crucial to detect those anomalies that will offer wrong information in such a way that the automated vehicle will take the wrong decision (e.g., collide with the pedestrian or cyclist approaching). As the overview in \cite{Khalid2020} shows, cyber-attacks can also induce other anomalies (e.g. in-vehicle communication), but the focus of this research is on attacks comings from infrastructure to the vehicle that can lead to unsafe situations. 
Therefore, this work focuses specifically  on \textit{instant}, \textit{bias}, and \textit{gradual drift} anomalies, claiming that e.g. the noise and burst anomalies can be classified as an intense temporal instant, bias and/or drift anomalies, and the \textit{miss} anomaly will be handled by the Target Tracker internally. These types of anomalies pose the highest threat and are the most dangerous for cooperative automated vehicles, \cite{Petit2014}\cite{Mo2010}. 
Since in the considered scenario the malicious data comes from a RSU, the focus in this work will be on detecting sensor anomalies (i.e. any sensor embedded in or attached to the vehicle, as well as sensors located in the environment).  Anomalies that come from the fused state output are not addressed here as this poses a different type of problem. 


\section{Problem Description}\label{sec:problemdescription}

While I2V can enhance the object detection capabilities of an AD vehicle in terms of field-of-view and sensor redundancy, communicating object information for sensor fusion does however introduce a cyber security threat as the data could potentially be spoofed or tampered with. Using this tampered data for sensor fusion could result in wrong object position or velocity estimations and as a result could lead to safety-critical situations.

When anomalies are detected in the fused state estimate (in the World Modelling block in Fig. \ref{safety}), a driver can be warned that the perception model might be compromised or the AD-system can bring the vehicle to a safe state. However, a compromised integrity of the state estimate cannot be avoided. By monitoring the sensors individually for anomalies, their health can be determined and compromised sensors can be temporarily ignored or permanently disabled in the sensor fusion algorithm. 

\subsection{System description}
The  automated vehicle considered is equipped with dedicated automotive sensors (radar, camera, lidar, GPS) as well as dedicated units for software and communication.
The RSU is a monitoring system at the side of the intersection, e.g. cameras which detect the VRUs. The communication is done via the Collective Perception Message (CPM) receiver and object parser which takes the relative camera object detections and transforms them to a global reference frame, so that they can be correlated with the vehicle's reference frame. This parser also adapts the information to an interface for the Target Tracker software.
\label{ScenarioDescription}
\subsection{The Scenario and its variations}
\begin{figure}[h!]
    \centering
    \includegraphics[clip, trim=0mm 00mm 20mm 30mm, width=0.95\columnwidth]{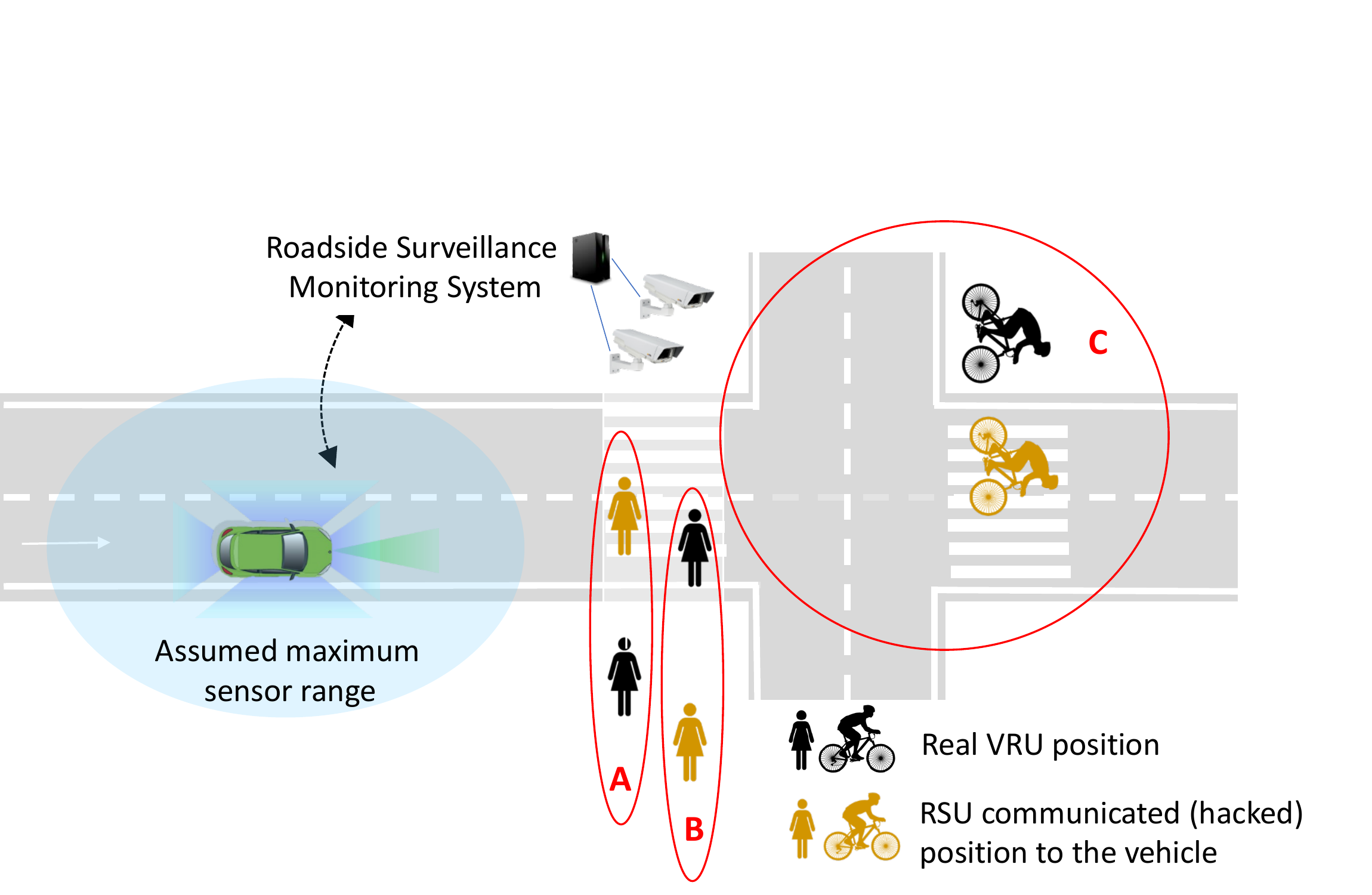}
    \caption{Examples of scenarios for anomalous RSU data}
    \label{fig:faults}
\end{figure}

In the considered scenario, an AD vehicle is approaching the intersection and receives malicious information from the hacked RSU, which is the only sensor considered in this study as affected by the attack. The vehicle detects the attack and mitigates it by ignoring the faulty external sensor information in the Target Tracker. 
In this scenario, a RSU transmits the wrong position of a VRU (a pedestrian or a cyclist), as shown in Fig. \ref{fig:faults}. 
Examples of variations include: \\
\textit{(i)} a VRU at the side of the intersection, waiting to cross. The hacked RSU communicates that the VRU is in the intersection instead (case A in Fig. \ref{fig:faults}), possibly causing the AD vehicle to unnecessarily stop at the crossing. This can lead, for example, to other dangerous situations for upcoming traffic behind the ego vehicle;\\
  \textit{(ii)} a VRU is crossing the intersection (case B in Fig. \ref{fig:faults}). The hacked RSU communicates that the VRU is located on the side of the intersection. The vehicle is intentionally not alerted about the passing VRU, which could lead to a collision;
\\
\textit{(iii)} A VRU is either at the side of the intersection or at different locations on the intersection. The RSU sends wrong information about the heading or velocity of the VRU to the automated vehicle which can mislead prediction or intention classification models of the vehicle.

\subsection{Simulation environment}
\begin{figure}[h!]
    \centering
    \includegraphics[clip, trim=0mm 0mm 0mm 0mm, width=0.95\columnwidth]{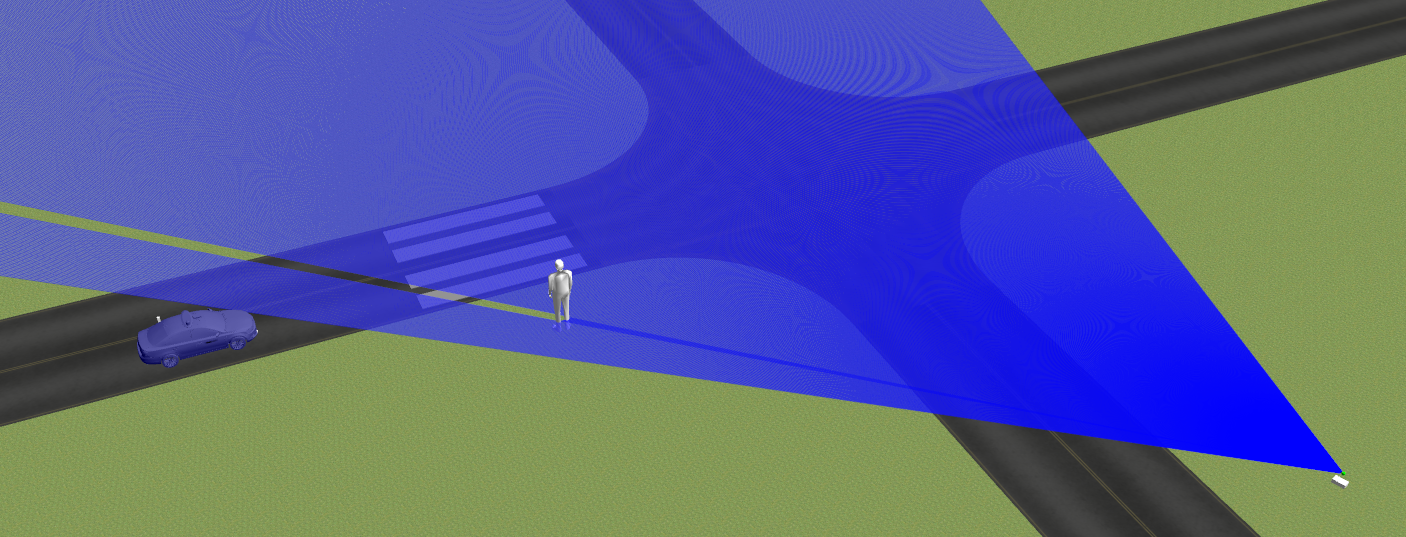}
    \caption{Gazebo-based simulation scenario}
    \label{fig:simulationscenario}
\end{figure}

Since real measured data is very hard to find for these scenarios, data is simulated for the considered scenario and its actors 
in a Gazebo simulator, \cite{Koenig2004} as depicted in Fig.  \ref{fig:simulationscenario}. The intersection's road network is defined using the OpenDrive standard, \cite{Dupuis2010}. The scenario itself, including actions and events, is described in the OpenScenario format \cite{OpenScenario}. 
To mimic realistic intrusion situations that can lead to unsafe situations, when the pedestrian is in the field of view of all on-board vehicle sensors as well as the RSU, an anomaly is injected as described next. Furthermore, to make the simulation as realistic as possible, sensor noises are introduced for both the on-board vehicle sensors, as well as the road side camera. Following typical sensor noises at automotive object tracking sensors, the noises levels, $\sigma$, are independent zero-mean white noises with the following standard deviations (used from~\cite{dePonteMuller2017} as $3\sigma$ values of a normal distribution): 
\begin{align}
    \sigma^{(1)}_{x}, \sigma^{(1)}_{y} &= 0.03 \ [m],\:
    &\sigma^{(2)}_{x}, \sigma^{(2)}_{y} &= 0.0067 \ [m],\nonumber\\
    \sigma^{(2)}_{v_x}, \sigma^{(2)}_{v_y} &= 0.17 \ [m/s], \:
    &\sigma^{(3)}_{x}, \sigma^{(3)}_{y} &= 0.03 \ [m], \nonumber\\
    \sigma^{(3)}_{v_x}, \sigma^{(3)}_{v_y} &= 0.17 \ [m/s], \:
    &\sigma^{(4)}_{x}, \sigma^{(4)}_{y} &= 0.03 \ [m], \nonumber\\
    \sigma^{(4)}_{v_x}, \sigma^{(4)}_{v_y} &= 0.17 \ [m/s], \: &\sigma^{(4)}_\theta,\sigma^{(4)}_{v_\theta}&=0.011 \ [rad] \label{eq:noisedefinition}
\end{align}
where the counter values $j\in \{1,2,3,4\}$ indicate the sensor source of the signals, being the radar, the lidar, the camera and the RSU respectively. For ease of notation, this counter $j$ is used throughout the rest of the paper with identical meaning. Moreover, the subscripts $x$, $y$, $v_x$, $v_y$ refer to the objects' longitudinal and lateral positions and velocity respectively. 

\subsection{Anomaly injection} \label{AnomalyInjection}
It is assumed that the vehicle is observed in 2D by the RSU, measuring   at each time sample $k$ the positions $x_{k}^{(4)},y_{k}^{(4)}$,  the heading $\theta_{k}^{(4)}$, and the corresponding velocities  $v_{x_k}^{(4)}$, $v_{y_k}^{(4)}$, $v_{\theta_k}^{(4)}$. The RSU measurement state is denoted by $u_k^{(4)}$ and defined as 
\begin{equation}
    u_k^{(4)} = \left[ \begin{array}{cccccc}
         x^{(4)}_k,& 
         y^{(4)}_k,& 
         \theta^{(4)}_k,&
         v_{x_k}^{(4)},& 
         v_{y_k}^{(4)},& 
         v_{\theta_k}^{(4)}
    \end{array}
    \right]^\intercal.
\end{equation}
The error injected for the different anomalies is defined by
\begin{equation}
    e_k =  \left[ \begin{array}{cccccc}
         e^x_k,& 
         e^y_k,&
         e^{\theta}_k,&
         e^{v_x}_k,&
         e^{v_y}_k,& 
         e^{v_\theta}_k
    \end{array}
    \right]^\intercal,
\end{equation}
with $(\cdot)^x$ referring to the anomaly injected in each respective dimension. The hacked output going into the vehicle on-board sensor fusion algorithm at time $k$ will be denoted by $\Hat{u}_k^{(4)}$.

Three different injected anomalies are defined: $e^D$ for gradual drift, $e^B$ for bias, and $e^I$ for instant. Usually, $||e^D|| < ||e^B|| < ||e^I||$, resulting in three different hacked outputs,
\begin{equation}
    \Hat{u}_k^{(4)} = u_k^{(4)}+e^I\Delta(k, k_{anomaly}), 
\end{equation}
for instant anomalies, with $\Delta(k, k_{anomaly})$ a Kronecker delta function at the anomaly time $k_{anomaly}$,
\begin{equation}
    \Hat{u}_k^{(4)} = u_k^{(4)} + e^B,
\end{equation}
for bias anomalies, and
\begin{equation}
    \Hat{u}_k^{(4)} = u_k^{(4)} + \Hat{u}_{k-1}^{(4)} - u_{k-1}^{(4)} + e^D,
\end{equation} 
for gradual drift anomalies. The \textit{drift} and \textit{bias} anomalies will be injected for different durations, from time $k_{start}$ until $k_{end}$, such that
\begin{align*}
    &e_k \neq 0 \quad for \quad k_{start} \leq k < k_{end}, \\   
    &e_k = 0 \quad for \quad k < k_{start} \quad || \quad  k \geq k_{end},
\end{align*}

where $k_{start}$ is chosen such that the pedestrian is observed by the RSU as well as by all on-board vehicle sensors at the time of the anomaly injection.

\section{In-vehicle cyber threats anomaly detection}\label{algorithm}
Using simulated data from Section \ref{sec:problemdescription}, the scenario depicted in   Fig.  \ref{fig:scene} and the available in-vehicle sensors and the RSU measurements, a fault diagnosis can be done. 

\subsection{Multi-sensor EKF anomaly detection algorithm}
\begin{figure}[t]
 \centering
    \includegraphics[clip, trim=70mm 40mm 104mm 75mm,width=0.9\columnwidth]{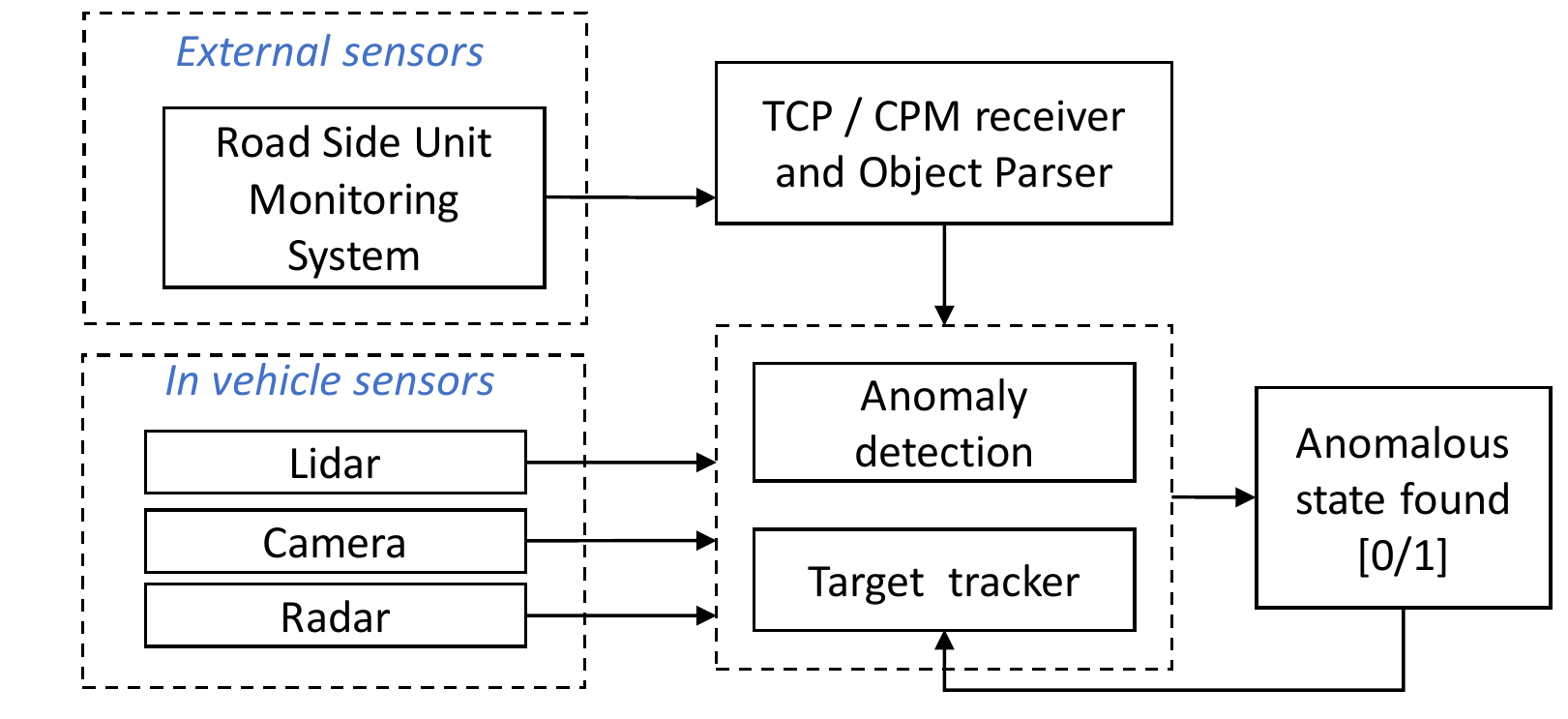}
    \caption{Schematic representation of the anomaly detection algorithm and how it interacts with the Target Tracker.}
    \label{fig:architecture}
\end{figure}

\begin{figure}[t]
    \centering
    \includegraphics[clip, trim=0mm 0mm 0mm 6mm,width=\columnwidth]{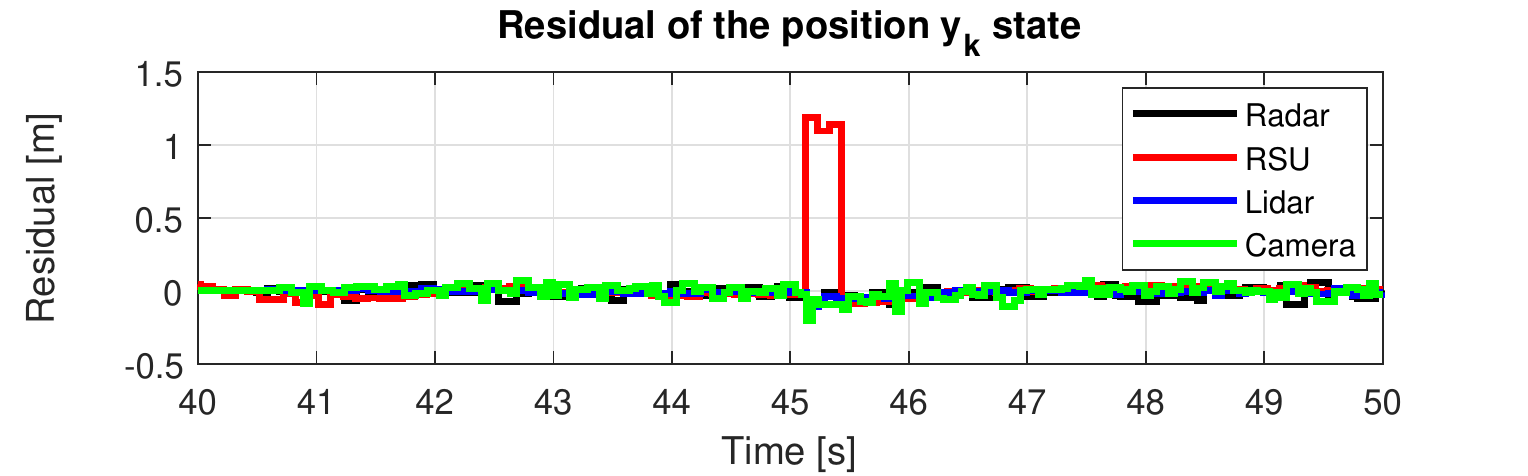}
    \caption{An example of the EKF residuals for a bias fault of $e^B_{x}=1.28$m introduced on the $y_k$ measurement of the RSU at time $t=45$ seconds.}
    \label{fig:residual}
\end{figure}

The algorithm for detecting anomalous sensor behaviour follows the architecture shown in Fig. \ref{safety} and the more detailed view is shown in Fig. \ref{fig:architecture}.  The anomaly detector is designed for detection of an anomaly on any of the sensor states. Once an anomaly is detected, the affected sensor state is disregarded in the Target Tracker for as long as the anomaly persists. This information can be used to disregard that measurement from the sensor fusion of the Target Tracker, in order to mitigate anomalies in the fused estimate of a certain state.
For detection of the aforementioned fault types, an Extended Kalman Filter (EKF) 
is combined with a $\mathcal{L}_2$ norm-based test for residual evaluation~\cite[Section 11.2.2]{Ding20081}. The EKF algorithm uses an omnidirectional motion model with state $X=\left[x,\:y,\:\theta,\:v,\:v_\theta,\:a\right]$ where $x,\:y$ denote the longitudinal and lateral positions and $\theta$ represents the heading of the target with respect to the automated vehicle. With $v$ and $a$ we denote the resultant planar velocity and acceleration and $v_\theta$ the first time derivative of the heading. These resultant components can be expressed in the velocities $v_x,\:v_y$ and accelerations $a_x,\:a_y$ as follows:
\begin{eqnarray*}
    v&=\sqrt{v_x^2+v_y^2}\\
    a&=\sqrt{a_x^2+a_y^2}
\end{eqnarray*}
In the model, a constant acceleration $a$ and a constant yaw-rate $v_\theta$, is assumed. This assumption is made due to the fact that we consider a VRU (e.g., a pedestrian or a cyclist) in the scenario and thus can assume low-dynamic behavior. The model of this VRU can be denoted as
 \begin{align*}
     x_{k+1}&=x_k+v_{k}\cos{(\theta_k)}\Delta_t+\frac{1}{2}a_{k}\cos{(\theta_k)}\Delta_t^2+w^{(1)}_k,\\
     y_{k+1}&=y_k+v_{k}\sin{(\theta_k)}\Delta_t + \frac{1}{2}a_{k}\sin{(\theta_k)}\Delta_t^2+w^{(2)}_k,\\
     \theta_{k+1}&=\theta_k+v_{\theta_k}\Delta_t+w_k^{(3)},\\
     v_{{k+1}}&=v_{k}+a_{k}\Delta_t+w_k^{(4)},\\
     v_{\theta_{k+1}}&=v_{\theta_k}+w_k^{(6)},\\
     a_{{k+1}}&=a_{k}+w_k^{(7)},
 \end{align*}
where $k$ denotes the discrete time counter, $w_k^{(i)}\sim(0,\sigma^2_i)\:\forall i$ is a zero-mean white process noise and $\Delta_t$ denotes the discrete sampling step. Note that the VRU is modelled as an undriven system, i.e., the object is only externally observed, using sensors on the automated vehicle as well as the RSU measurements. These available measurements can be modelled in the VRU model, used for EKF synthesis, as
\begin{align*}
    u_k&=h(X_k)+z_k,
\end{align*} 
where $z_k$ is a vector signal of independent zero-mean white noises and the measurement matrix $H$ is defined by
\begin{align}
    h^{(1)}(X_k) &= \delta_k^{(1)}\left[x_k,\:y_k\right]^\intercal,\nonumber\\
    h^{(2)}(X_k) &= \delta_k^{(2)}\begin{bmatrix}x_k, y_k, v_k\cos(\theta_k), v_k\sin(\theta_k)\end{bmatrix}^\intercal,\nonumber\\
    h^{(3)}(X_k) &= \delta_k^{(3)}\begin{bmatrix}x_k, y_k, v_k\cos(\theta_k), v_k\sin(\theta_k)\end{bmatrix}^\intercal,\nonumber\\
    h^{(4)}(X_k) &=\delta_k^{(4)} \begin{bmatrix}x_k, y_k, \theta_k,
    v_k\cos(\theta_k), v_k\sin(\theta_k),v_\theta\end{bmatrix}^\intercal,\nonumber
    \end{align}
where $\delta_k^{(j)}\forall j$ denotes a set of indicator signals, which depend on the availability of a new measurement $h_k^{(j)}\forall j$. I.e., when a specific sensor measurement is available, the indicator function for that measurement is set to $1$, otherwise it is equal to $0$. The fused state prediction $\hat{X}_{k|k-1}$ of the EKF and its covariance matrix can be denoted as 
\begin{align}
    \hat{X}_{k|k-1}&=f(\hat{X}_{k-1|k-1}),\nonumber\\
    P_{k|k-1} &= A_kP_{k-1|k-1}A_k^\intercal+Q,\label{eq:processcovariance}\\
    A_k&=\left.\frac{\partial f(X_k)}{\partial X_k}\right|_{\hat{X}_{k-1|k-1}},\nonumber
\end{align}
where $P_{k,k-1}$ is the state prediction covariance matrix, $A_k$ is the jacobian of $f(X_k)$ and $Q$ is the process noise covariance matrix. The Kalman gain, used to update the internal state estimate for the current time-step, is calculated by 
\begin{align}
    H_k&=\left.\frac{\partial h^{(j)}}{\partial X_k}\right|_{\hat{X}_{k|k-1}},\nonumber\\
    S_{k} &= \varepsilon_k^{(j)} H_kP_{k|k-1}H_k^\intercal\varepsilon_k^{(j)}+R^{(j)},\label{eq:faultreaction I}\\
    K_{k} &= P_{k|k-1}H_k^\intercal\varepsilon_k^{(j)} S_{k}^{-1}\label{eq:faultreaction II},
\end{align}
where $H_k$ represents the linearized state measurement matrix, $R$ represents the sensor noise covariance matrix for each respective sensor $j$ and $K_k$ represents the Kalman gain. Finally, $\varepsilon_k^{(j)}$ is a set of indicator signals dependent on the presence of a fault for a certain sensor measurement, something which is explained further on in the residual evaluation. For the next time-step, the generated residual, the state estimate and  its covariance matrix are predicted by the following equations
\begin{align}
    r_k&=(u_k-h^{(j)}(\hat{X}_{k|k-1}))\label{eq:innovation}\\
    \hat{X}_{k|k}&=\hat{X}_{k|k-1}+K_k r_k,\label{eq:updatestep1}\\
    P_{k|k} &= \left(I-K_kH_k\right)P_{k|k-1}.\label{eq:updatestep2}
\end{align}
where $r_k$ denotes the innovation residual signal. Note that this residual $r_k$ is used as a fault detection indicator, i.e., the residual signal is used to detect whether a measurement deviates too much from the predicted measurement. This concludes the EKF algorithm. An illustrative example of the residuals coming out of the algorithm is given in Fig. \ref{fig:residual}. Herein, a bias anomaly is injected in the RSU and it can be observed that the RSU residual is sensitive to this bias.
As mentioned before, the EKF generates a residual  signal~\eqref{eq:innovation} which gives information about the deviation of the  predicted measurement, calculated based on the predicted state $\hat{X}_{k|k-1}$ and the measurement model with the measurements $u_k$. A next layer of residual evaluation is needed. The residual evaluation is done using a $\mathcal{L}_2$ norm-based detector of the form
\begin{align}
    \hat{r}_k = \frac{1}{n}\sum_{i=k-n}^{k} r_i\circ r_i,\label{eq:movingavg}
\end{align}
where the operator $\circ$ represents the pointwise Hadamard product. Each entry of the resulting vector of signals $\hat{r}_k$ is compared with a pre-defined positive-valued threshold vector $\alpha^{(j)}$ of which the size is equal to the number of measured states per measurement of each respective sensor $j$. Subsequently, given a predetermined time horizon $n$, an appropriate reaction is given using the indicator function $\varepsilon_k$, 
\begin{align}
   \hat{r}_{i,k}^{(j)}&\leq \alpha^{(j)}_{i}\xrightarrow[]{\text{No fault detected}}\varepsilon_{i,k}^{(j)}=1,\nonumber\\
   \hat{r}_{i,k}^{(j)}&> \alpha_{i}^{(j)}\xrightarrow[]{\text{Fault detected}}\varepsilon_{i,k}^{(j)}=0,\quad\forall i,j,\label{eq:threshold}
\end{align}
where $i$ selects the index of the measurement state. The reaction of a detected fault appears directly in the equations~\eqref{eq:faultreaction I} and~\eqref{eq:faultreaction II}, where the affected measurement state is no longer taken into account in the update step~\eqref{eq:updatestep1},~\eqref{eq:updatestep2}. However, the affected measurement is taken into account in ~\eqref{eq:innovation} as to detect whether a fault is still present or not.

\section{Results Analysis}\label{results}
As mentioned in Section \ref{sec:problemdescription}, in order to benchmark the performance of the anomaly detection algorithm, simulations are carried out with anomalies of varying size and duration. 
For the \textit{instant} anomaly, an anomaly size $e^I_x$ is chosen following a logarithmic distribution of 10 samples between \textit{0.1\:m} and \textit{10\:m}. The duration of the instant anomaly is $d=0.05\:s$. This results in a total of 10 tests for evaluation of the algorithm in the case of instant anomalies.

For the \textit{bias} anomaly, an anomaly size $e^B_x$ is chosen following a logarithmic distribution of 5 samples between \textit{0.1\:m} and \textit{3\:m}. A logarithmic distribution is chosen to determine which order of magnitude for fault size can be flagged by the anomaly detector. Furthermore, the anomaly duration is varied with the following variations: $d=[0.25, 0.5, 1.0, 2.5]\:s$. This results in a total of 20 tests for evaluation of the algorithm in the case of bias faults.
For the \textit{drift} anomaly, an anomaly size $e^D_x$ is chosen again by following a logarithmic distribution of 5 samples between \textit{0.1\:m/s} and \textit{3\:m/s}. Furthermore, the anomaly duration is varied again with the following variations: $d=[0.25, 0.5, 1.0, 2.5]\:s$. This results in a total of 20 tests for evaluation of the algorithm in the case of drift faults.


\subsection{Statistical results}\label{sec:statisticalresults}
\begin{figure}[h!]
    \centering
    \includegraphics[width=\columnwidth]{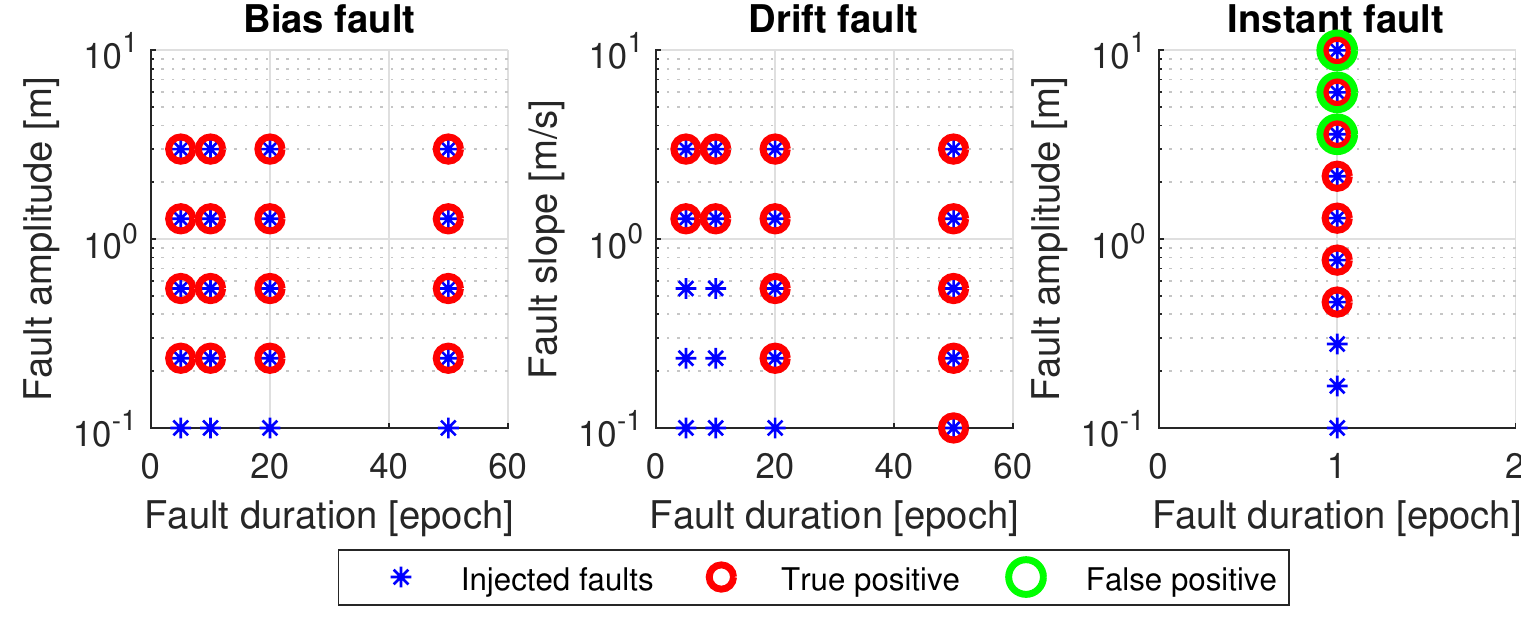}
    \caption{Results of the fault detection performance and false positive robustness, $72\%$ true positive, $6\%$ false positive}
    \label{fig:Statistical_res}
\end{figure}

To test the performance and reliability of the fault detection algorithm, the tests introduced before are applied online in the simulation environment. The initial state covariance matrix and state are chosen as $P_0=I,\:\hat{X}_0=0$. The measurement noise covariance matrix $R$ from~\eqref{eq:faultreaction I} is chosen by directly substituting squared of the measurement noises defined in~\eqref{eq:noisedefinition}. The process noise covariance matrix, $Q$, from~\eqref{eq:processcovariance} is chosen as $Q=0.001\cdot I$. The residual evaluation horizon~\eqref{eq:movingavg} is chosen as $n=30$ and finally the evaluation thresholds~\eqref{eq:threshold} are chosen to be equal to $\alpha_{x}=\:\alpha_y=\:\alpha_\theta=0.18$ for all position states and $\alpha_{v_x}=\:\alpha_{v_y}=\:\alpha_{v_\theta}=0.7$ for all velocity states. These threshold have been determined iteratively using the magnitude of the noise as a starting point. Using these parameters and initial conditions, the results are given in Fig. \ref{fig:Statistical_res}. This type of figure is interpreted as follows; faults are injected of certain type (bias, drift or instant) with a certain magnitude (y-axis) and a certain duration (x-axis). If the injected fault (blue star) is detected, it is marked (red circle). If a fault is detected on a sensor for which no fault is injected, it is marked as false positive (green circle). Using this graphical representation, it becomes straightforward to observe the actual fault, whether it was detected or not and whether any false positives on other sensors occurred.

These results show that the fault detection algorithm is able to detect $72\%$ of the injected faults, where in $6\%$ of the test cases a false positive is also detected on a different sensor measurement state. The undetected faults either have a too low maximum value (effect of threshold $\alpha$, preventing noise on the residual to be classified as a fault), or the fault duration is too short (filtering effect of the fault horizon $n$). The false positives that are detected at high amplitudes of faults are caused by the combined internal estimate, $\hat{X}_k$, being pushed up due to the faulty measurement before it is being disregarded. This phenomenon causes a discrepancy between the internal estimate and the healthy sensors, hence potentially classifying them as faulty.
\subsection{Case-study when losing sensor redundancy}
The results from Section~\ref{sec:statisticalresults} show that the algorithm is able to detect faults above a certain uncertainty threshold very well. This threshold, determined by $\alpha$, could be set lower, yet that is not desirable since it would compromise robustness as the remaining noise in the residual could induce false positives. However, it also shows that the algorithm starts detecting false positives for high fault amplitudes. The faulty measurements are not detected instantaneously due to the filtering effect of the EKF as well as the estimation horizon $n$ of the residual evaluation. Therefore, the faulty measurement is used for a certain time to update the internal state estimate. 

For a sufficiently high fault, the state estimate temporarily increases up to such an extent, that non-faulty measurements are being considered faulty due to their difference with the internal state estimate. It is expected that this effect would grow even further when losing sensor redundancy. In order to test this hypothesis, the matrix is retested without the lidar and radar sensor measurements, using the same initial conditions and parameters as in Section~\ref{sec:statisticalresults}. 
\begin{figure}[h!]
    \centering
    \includegraphics[width=\columnwidth]{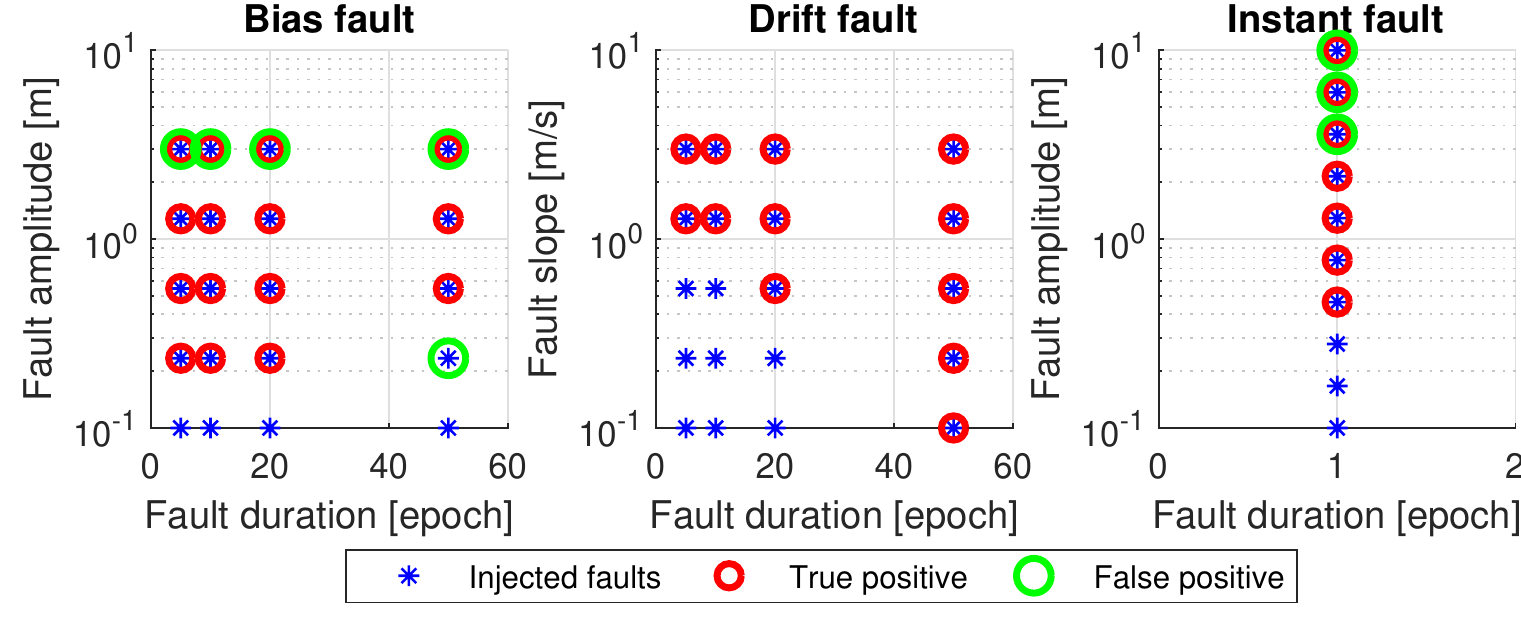}
    \caption{Results of the fault detection performance and reliability without radar and lidar redundancy, $68\%$ true positive, $22\%$ false positive}
    \label{fig:Statistical_res_red}
\end{figure}
 The results of this analysis are shown in Fig. \ref{fig:Statistical_res_red}. Here, one can observe that the cumulative sum of the false positive detection increases. This can be explained by the fact that there are only two sensors that construct the internal state estimate, $\hat{X}_k$, and as such the error between $\hat{X}_k$ and the healthy measurements increase as well. In comparison with the full redundancy-case however, the number of true positives only decreased by $4\%$, showing robustness of the algorithm for true positive detection in the case of loss of redundancy. 
\subsection{Discussion}
Detecting anomalies from external sensors as defined here does not offer information nor does it determine the cause of the fault (i.e., fault classification). The primary interest here was to detect the fault and as a mitigating action, to no longer incorporate the affected state in the sensor fusion. It depends on the safety criticality of the function whether the driver of the vehicle should be made aware of the cause of a fault.

The external observed object, is only observed using the vehicle's sensors and sensors from the road infrastructure, while its intention and actual movement are unknown (no sensors are present on this external object). 
This implies that, to have a reliable fault detection algorithm, a degree of sensor redundancy is required which is a limitation of this algorithm. By using all these sensors to predict the same internal state of the externally observed object, the internal state estimate can be pushed off its true trajectory by a fault with sufficient amplitude. Nonetheless, it is possible that this problem can be circumvented by augmenting the model with predicted or known intentions of the externally observed object. 

\section{Conclusions and future work}\label{conc}
The contribution of this work is twofold. The anomaly detection algorithm introduced here provides insights into the possibility to detect different types of attacks from the automated vehicle's point of view, in I2V-enabled road intersections. The results show that majority of relevant and potentially dangerous anomalies can be detected,  especially if the fault magnitude is limited and in-vehicle sensor redundancy exists. In addition, the results show the benefits of having sensor redundancy for object tracking and the limitation when this does not exist. This implies the automated vehicle can maintain a safe and comfortable operation even in the face of I2V cyber-attacks.
This methodology, used to detect attacks in the I2V communication can, be applied on detecting in-vehicle sensor anomalies as well, using a fusion of in-vehicle and I2V sensor data.

Future work includes the extension towards more scenarios, investigating robustness to network effects (delays, package drop), incorporating more I2V signals (e.g. traffic light information) and incorporating the EKF anomaly detection method with the vehicle control mitigation strategies for cases where the anomalies can be detected. Furthermore, the extension of the method with data-driven methods can be considered to improve the false positive and false negative rates in case of large faults amplitudes. 

\section{Acknowledgments}
This work is supported by EU Horizon 2020 R{\&}D programme under grant agreement No. 783119, project SECREDAS (Product Security for Cross Domain Reliable Dependable Automated Systems).
\printbibliography

@inproceedings{Park2015,
author = {Park, J. and Ivanov, R. and Weimer, J. and Pajic, M. and Lee, I.},
title = {Sensor Attack Detection in the Presence of Transient Faults},
year = {2015},
booktitle = {Proceedings of the ACM/IEEE Int. Conference on Cyber-Physical Systems},
pages = {1–10},
numpages = {10},
}

@ARTICLE{van_Wyk2019,
author={F. {van Wyk} and Y. {Wang} and A. {Khojandi} and N. {Masoud}},
journal={IEEE Transactions on Intelligent Transportation Systems},
title={Real-Time Sensor Anomaly Detection and Identification in Automated Vehicles},
year={2019},
pages={1-13},
}

@ARTICLE{Bhuyan2014,
author={M.H. {Bhuyan} and D.K. {Bhattacharyya} and J.K. {Kalita}},
journal={IEEE Communications Surveys Tutorials},
title={Network Anomaly Detection: Methods, Systems and Tools},
year={2014},
volume={16},
number={1},
pages={303-336},
}

@article{Sharma2010,
author = {Sharma, A. and Golubchik, L. and Govindan, R.},
year = {2010},
month = {01},
pages = {},
title = {Sensor Faults Detection Methods and Prevalence in Real-World Datasets},
volume = {6},
journal = {TOSN}
}

@article{Grembek2019,
author = {Grembek, O. and Kurzhanskiy, A. and Medury, A. and Varaiya, P. and Yu, M.},
year = {2019},
month = {05},
pages = {396-410},
title = {Making intersections safer with I2V communication},
volume = {102},
journal = {Elsevier: Transportation Research Part C: Emerging Technologies}
}

@BOOK{Ding20081,
author={Ding, S.X.},
title={Model-based fault diagnosis techniques: Design schemes, algorithms, and tools},
year={2008},
pages={1-473},
publisher={Springer Berlin Heidelberg},
}

@article{Wang2019,
author = {Wang, Y. and Masoud, N. and Khojandi, A.},
year = {2019},
month = {11},
title = {Real-Time Sensor Anomaly Detection and Recovery in Connected Automated Vehicle Sensors},
journal = {IEEE transactions on intelligent transportation systems}
}

@article{Petit2014,
author = {Petit, J. and Shladover, S.},
year = {2014},
month = {09},
title = {Potential Cyberattacks on Automated Vehicles},
journal = {IEEE transactions on intelligent transportation systems}
}

@article{Mo2010,
author = {Mo, Y. and Garone, E. and Casavola, A. and Sinopoli, B.},
year = {2010},
month = {12},
title = {False Data Injection Attacks against State Estimation in Wireless Sensor Networks},
journal = {IEEE conference on Decision and Control}
}

@article{Dupuis2010,
author = {Dupuis, M. and Strobl, M. and Grezlikowski, H.},
year = {2010},
title = {OpenDRIVE 2010 and beyond – status and future of the de facto standard for the description of road networks},
journal = {Driving Simulation Conference Europe},
pages={231-242}}

@article{Koenig2004,
author = {Koenig, N. and Howard, A.},
year = {2004},
month = {09},
title = {Design and use Paradigms for Gazebo, An Open-Source Multi-Robot Simulator},
journal = {IEEE/RSJ International Conference on Intelligent Robots and Systems}
}

@INPROCEEDINGS{Arashsafety,
  author={Y. {Luo} and A.K. {Saberi} and T. {Bijlsma} and J.J. {Lukkien} and M. {van den Brand}},
  booktitle={IEEE International Systems Conference}, 
  title={An architecture pattern for safety critical automated driving applications: Design and analysis}, 
  year={2017},
  pages={1-7},
}

@phdthesis{arashphd,
title = "Functional Safety: A New Architectural Perspective: Model-Based Safety Engineering for Automated Driving Systems",
author = "{Khabbaz Saberi}, Arash",
year = "2020",
publisher = "Eindhoven University of Technology",
}

@inproceedings{Checkoway2011,
  title={Comprehensive Experimental Analyses of Automotive Attack Surfaces},
  author={Stephen Checkoway and D. McCoy and B. Kantor and Danny Anderson and H. Shacham and S. Savage and Karl Koscher and Alexei Czeskis and F. Roesner and T. Kohno},
  booktitle={USENIX Security Symposium},
  year={2011}
}

@ARTICLE{Ellen2012,
  author={E. {van Nunen} and M.R.J.A.E. {Kwakkernaat} and J. {Ploeg} and B.D. {Netten}},
  journal={IEEE Transactions on Intelligent Transportation Systems}, 
  title={Cooperative Competition for Future Mobility}, 
  year={2012},
  volume={13},
  number={3},
  pages={1018-1025},
}

@INPROCEEDINGS{Panda2018,
author = {Panda, S. and Oliver, I. and Holtmanns, S.},
year = {2018},
month = {06},
pages = {119-126},
booktitle={Asian Control Conference}, 
title = {Behavioural modelling of attackers’ choices},
}

@misc{iso,
author = {ISO 26262-1},
year = {2011},
title = {Road vehicles - functional safety. {G}eneva, {S}witzerland: International Organization for Standardization},
}

@misc{OpenScenario,
author = {ASAM},
year = {2020},
  title = {ASAM OpenSCENARIO V1.x},
  note= {https://www.asam.net/project-detail/asam-openscenario-v1x/ (Accessed: 2020-10-23)}
}

@INPROCEEDINGS{Lampiri2017,
  author={E. {Lampiri}},
  booktitle={Asian Control Conference}, 
  title={Sensor anomaly detection and recovery in a nonlinear autonomous ground vehicle model}, 
  year={2017},
  pages={430-435},
}

@INPROCEEDINGS{negi2020,
  author={N. {Negi} and O. {Jelassi} and H. {Chaouchi} and S. {Clemençon}},
  booktitle={International Conference on Artificial Intelligence in Information and Communication}, 
  title={Distributed online Data Anomaly Detection for connected vehicles}, 
  year={2020},
  volume={},
  number={},
  pages={494-500},
}

@ARTICLE{dePonteMuller2017,
author={de Ponte Müller, F.},
title={Survey on ranging sensors and cooperative techniques for relative positioning of vehicles},
journal={Sensors},
year={2017},
volume={17},
number={2},
art_number={271},
document_type={Article},
}

@article{Khalid2020,
title = {Cyber-attacks in the next-generation cars, mitigation techniques, anticipated readiness and future directions},
journal = {Accident Analysis \& Prevention},
volume = {148},
pages = {105837},
year = {2020},
issn = {0001-4575},
author = {S. Khalid Khan and N. Shiwakoti and P. Stasinopoulos and Y. Chen},
}
\end{document}